\documentclass[10pt, letterpaper]{article}
\usepackage{graphicx}
\usepackage[margin=1in]{geometry}
\usepackage{hyperref}
\usepackage{amsmath}
\usepackage{amssymb}
\usepackage{newtxtext}
\usepackage{newtxmath}
\usepackage{svg}
\title{Parametric amplification in a Kerr Oscillator based on Ne FIB Nanobridges}
\usepackage{authblk}

\author[1,2]{Parth Bhandari%
\thanks{Corresponding author: 
\href{mailto:parth.bhandari.22@ucl.ac.uk}{parth.bhandari.22@ucl.ac.uk}}}
\author[2]{Laith Meti}
\author[2]{Gemma Chapman}
\author[1]{Edward Romans}
\author[2]{John Gallop}
\author[2]{Ling Hao}
\affil[1]{Department of Electronic and Electrical Engineering, University College London, London, UK}
\affil[2]{National Physical Laboratory, Teddington, UK}
\date{}
\begin{document}
\maketitle
\begin{abstract}
Superconducting circuits play a crucial role in the advancement of quantum computing and quantum sensing. Typically such circuits require the presence of a non-linear element, where the engineered anharmonicity (Kerr factor) and resonant linewidth determine the potential applications of the circuit. In this work we have fabricated Nb-based CPW resonators embedded with a DC SQUID incorporating Nb nanobridges as the weak links. We use two-tone spectroscopy to study the non-linear behaviour of the device at 15$\,$mK up to a field of 2.48$\,$mT. Under the application of a blue-detuned pump, the device shows a decrease in the resonant frequency which is used to estimate the Kerr factor. We further apply a red-detuned pump to go beyond the bifurcation threshold and observe the appearance of an additional idler mode with net gain. The gain of the device was maximized by further decreasing the pump frequency, showing a maximum amplification of 15$\,$dB. Finally, we show agreement between the Kerr non-linear oscillator model and the measured transmission spectrum, and highlight further design modifications to improve the gain and bandwidth of such devices.
\end{abstract}
\section{Introduction}
\label{sec:Introduction}
Nonlinear resonator circuits based on Josephson junctions (JJs) are central to microwave quantum technologies, where the nonlinear Josephson inductance (\textit{L}${\rm_J} = \hbar/(2eI{\rm_c} \rm{cos}(\phi))$), enables their use as mixers \cite{Taur1974,Taur1980}, amplifiers \cite{castellanos2007jpa,macklin2015twpa}, frequency combs \cite{lu2021comb}, and qubits \cite{krantz2019guide,blais2021cqed}. The ratio of the Kerr factor $K$ to the resonator linewidth $\kappa$ is used to classify potential applications of a JJ-based resonator \cite{Bourassa2012}. Such a resonator can be described by the Hamiltonian \cite{yurkebuks2006}
\begin{equation}
        H_r = \hbar \omega_0 A^{\dagger}A + \frac{\hbar}{2}K A^{\dagger}A^{\dagger}A A
\end{equation}
where $A$ and $A^{\dagger}$ are photon annihilation and creation operators respectively, and $\omega_0$ is the low-power angular resonant frequency. The Kerr factor is determined by the nonlinearity of the Josephson element and its participation in the resonant mode \cite{minev2021energy}. The total linewidth  $\kappa = \kappa_{\mathrm{int}} + \kappa_{\mathrm{ext}}$ is a combination of internal linewidth $\kappa_{\mathrm{int}}$ and external linewidth $\kappa_{\mathrm{ext}}$. The internal linewidth is set by two-level-system (TLS) and radiation losses which are impacted by both material quality and the resonator geometry. The external linewidth is set by the coupling capacitance. When $|K|/\kappa \gg 1$, the anharmonicity is resolved at the single-photon level and the circuit behaves as an artificial atom: an addressable two-level system, or qubit. This is the regime underlying circuit quantum electrodynamics \cite{blais2021cqed}, in which such artificial atoms have been used to synthesize nonclassical states of microwave light, such as Fock states \cite{hofheinz2008fock}, and to realize single-atom lasing \cite{astafiev2007lasing}. When $|K|/\kappa < 1$, the circuit is a weakly nonlinear resonator. This regime underlies the dispersive readout of superconducting qubits \cite{lin2014phaselocked}, the Josephson bifurcation amplifier \cite{vijay2009jba}, and the Josephson parametric amplifier (JPA) where a broad linewidth and strong overcoupling optimize the gain-bandwidth product and enable near-quantum-limited amplification \cite{castellanos2007jpa,  macklin2015twpa,yamamoto2008fluxdriven} and squeezing \cite{castellanos2008squeezing}. Within this weakly nonlinear family, devices can be further distinguished by their quality factor: most parametric amplifiers are low-$Q$ resonator, whereas the same parametric effects can equally be studied in a high-$Q$ resonator, as in ref. \cite{fanisani2021levelattraction} and the present work. A nonlinear resonator will mix through four-wave mixing, and under strong red-detuned driving will show a decrease in resonant frequency per added photon. This can accumulate enough intracavity photons to carry the cavity across its bifurcation threshold. Recent work has shown that for a high-$Q$ Al resonator with nanobridges as the Josephson element, this generates an idler mode with net output gain \cite{fanisani2021levelattraction}. \\
However, the use of Al has limitations because of its low critical field (10$\,$mT) giving it poorer tolerance to stray or applied fields \cite{murray2021material} in certain applications. For instance high-field quantum-sensing experiments, such as axion haloscopes, benefit from positioning the first-stage amplifier close to the detector to minimise microwave loss before amplification. However, this can expose the amplifier to residual fringe fields on the millitesla scale as reported in \cite{braggio2022haloscope,kim2023axion}. Conventional Al resonators are typically operated in perpendicular fields below approximately 100$\,\mu$T \cite{Song2009,Nsanzineza2014}. Operation at substantially higher fields has therefore required additional engineering, including the use of granular aluminium films \cite{Borisov2020} or narrower resonator geometries \cite{Samkharadze2016,Roy2025}. This constraint motivates the use Nb which has a higher critical field of 0.4$\,$T. In addition to its higher critical field, the higher critical temperature and superconducting gap of Nb reduce quasiparticle losses \cite{Anferov2024a}. \\
In the present work, we investigate a weakly nonlinear ($|K|/\kappa < 1$) quarter-wave CPW resonator terminated by a DC nanoSQUID incorporating Nb nanobridges fabricated by Ne Focussed Ion Beam (FIB) lithography. The use of nanobridges rather than tunnel junctions minimises the effective area presented to in-plane magnetic fields and thus maximises the device's in-plane field tolerance \cite{janssen2024magnetic}. In addition, nanobridges can be patterned monolithically in a single superconducting layer, avoiding the need for multilayer fabrication. The use of Ne FIB ensures that the Nb nanobridges are sufficiently short ($\sim$50$\,$nm) compared with the Ginzburg--Landau coherence length at mK temperatures to exhibit a Josephson-like current--phase relation \cite{likharev1979weaklinks,potter2025resonator}.  Incorporating two Nb nanobridges in a DC SQUID provides a flux-tunable Kerr nonlinearity \cite{bothner2024niobium,bhandari2026nbnanobridge}.\\
Previous work in the field used two-tone spectroscopy to determine the Kerr factor of Nb nanobridges embedded in a lumped-element \textit{LC} circuit at 2.4–3.2$\,$K \cite{bothner2024niobium}, and we presented preliminary measurements showing a nonlinear microwave response of a DC nanoSQUID-terminated CPW resonator at 15 mK \cite{bhandari2026nbnanobridge}. In the present work, we expand the preliminary study by systematically characterising the nonlinear response using a weak probe and a strong blue-detuned pump to extract the Kerr factor as a function of applied magnetic field at 15$\,$mT. We show the Kerr response persists up to an applied perpendicular field of 2.48$\,$mT, approximately an order of magnitude higher than the magnetic fields typically tolerated by conventional Al resonators. In addition, we investigate the red-detuned pumping regime, which drives the resonator towards the bifurcation regime where the four-wave-mixing parametric gain can develop. Our results show that above a well-defined pump-power threshold, the two-tone response develops a characteristic dip-and-peak structure associated with the emergence of an idler mode and net gain. By progressively lowering the pump frequency, and hence increasing its red detuning, we map the frequency-locking behaviour of the signal and idler modes and identify the pump condition that maximises the output gain. Under the optimised pump condition, we demonstrate, to our knowledge, the first four-wave-mixing parametric gain with Nb nanobridges. The measured responses are quantitatively reproduced by the established theoretical framework for driven Kerr nonlinear oscillators \cite{yurkebuks2006,bothner2024niobium}, providing a basis for further optimisation of Nb nanobridge-based parametric amplifiers.
\section{Device and Setup}
\label{sec:Device and Setup}
\begin{figure}[h]
\centering
\includegraphics[width=\columnwidth]{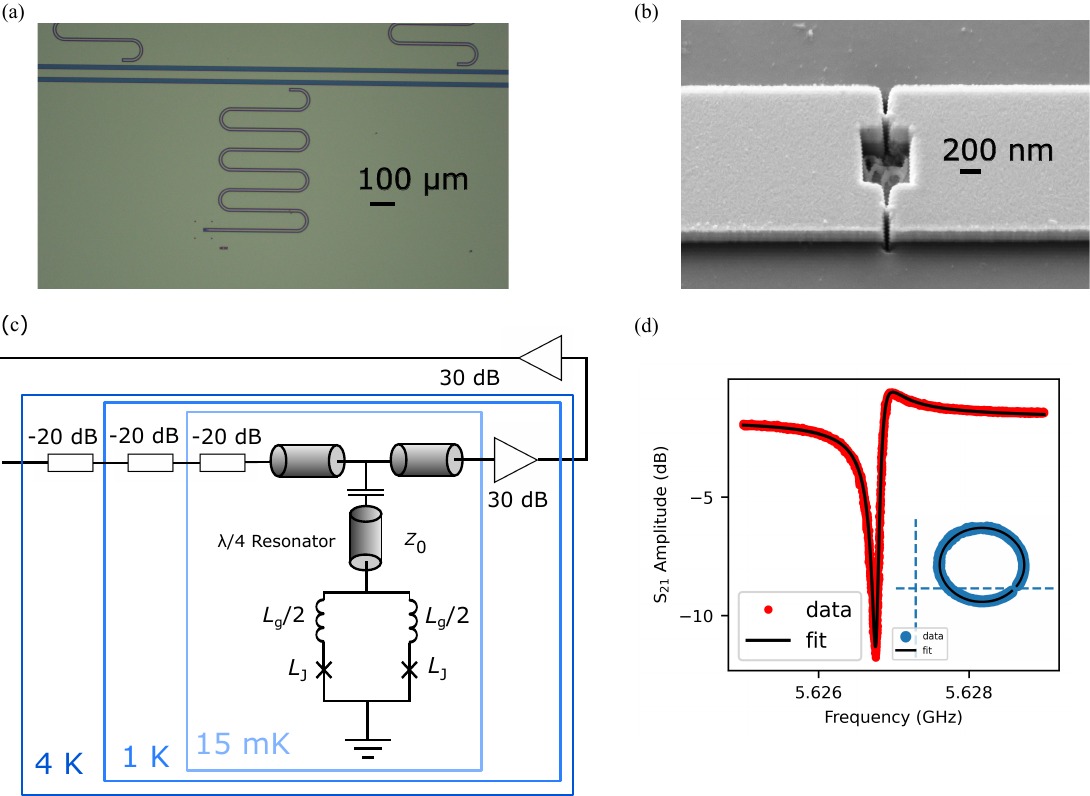}
\caption{(a) Optical image of a subsection of the chip showing three out of six resonators coupled to the central feedline. (b) SEM image of a Ne FIB DC SQUID embedded at the end of a resonator. (c) Schematic of the measurement setup showing attenuation and amplification at different stages. $L{\rm_g}$ is the total geometric loop inductance of the SQUID and $L\rm_{J}$ is the Josephson inductance of each nanobridge. (d) Plot showing the resonator transmission spectrum for $P\rm_{pr}  = $-105$\,$dBm with the data shown in red and the fit shown in black. The inset shows the circle fit in the complex plane, with the data shown in blue and the fit shown in black. The fit corresponds to a resonant frequency of 5.6268$\,$GHz, an internal quality factor, $Q\rm_{int} = 7.4\times10^4$, and an external quality factor, $Q\rm_{ext} = 4.5\times10^4$.}
\label{fig:fab}
\end{figure}

Our device was fabricated from a 150$\,$nm-thick Nb film on a 10$\,$mm$\times$10$\,$mm silicon substrate. The device has a central feedline that capacitively couples to six $\lambda/4$ resonators. We will discuss the non-linear behaviour of one of these resonators terminated by a DC nanoSQUID with resonant frequency of 5.6268$\,$GHz. The fabrication used Electron Beam Lithography (EBL) and Reactive Ion Etching (RIE) for defining the RF pads, Coplanar Waveguide (CPW) feedline, and CPW resonators. Subsequently a Ne Focused Ion Beam (FIB) was utilised to define the DC nanoSQUID and nanobridges. An optical image of a section of the chip showing three resonators coupled to the feedline is shown in Fig. \ref{fig:fab}(a). The end of the resonator was narrowed during the EBL process, and an SEM image of a subsequently incorporated Ne FIB nanoSQUID is shown in Fig. \ref{fig:fab}(b).  We have provided a detailed description of the fabrication process in \cite{bhandari2026nbnanobridge}.\\
We use a Bluefors LD dilution fridge fitted with a 2$\,$T superconducting magnet to cool down the chip to 15$\,$mK and study it under an applied magnetic field. The measurement setup is shown in Fig. \ref{fig:fab}(c), with 60$\,$dB attenuation on the input line, and 60$\,$dB amplification on the output line from a combination of a 4$\,$K HEMT amplifier and a Low Noise Amplifier (LNA) at room temperature. In the measurement protocol, we first studied the chip using a Rohde and Schwarz Vector Network Analyzer (VNA) to find the resonant frequency and quality factors using the circle fit method \cite{probst2015resonator} as shown in Fig. \ref{fig:fab}(d). We later added a strong drive (pump tone) using a Keysight signal generator. The two signals were combined using a power combiner at room temperature.\\
\section{Results}
\subsection{Blue-detuned Regime}
\label{sec:Blue-detuned Regime}
\begin{figure}[h]
    \centering
\includegraphics[width=\columnwidth]{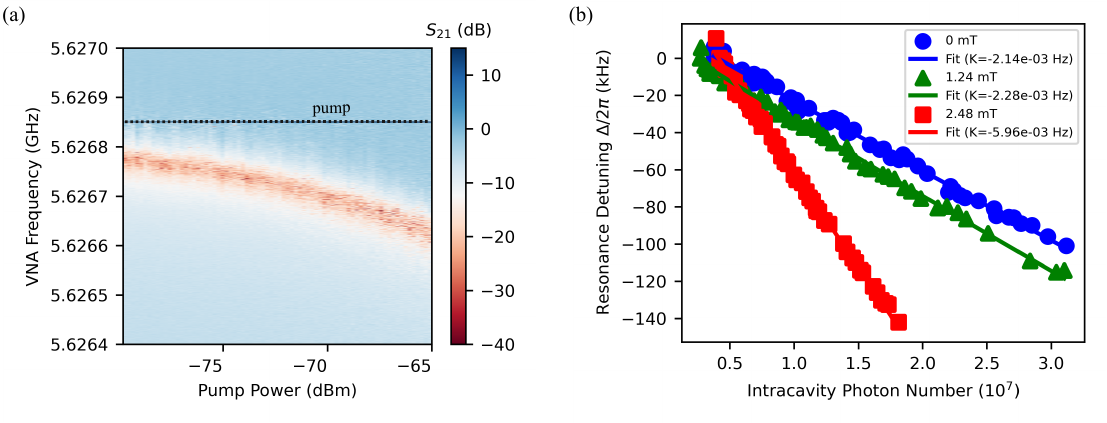}
    \caption{(a) A colour plot showing the transmission spectrum (S$\rm_{21}$) obtained for resonator with increasing pump power. The pump is annotated with a dotted black line, and the dip represented in red shows the shift in the resonant frequency of the circuit. (b) Plot showing the detuning of the resonant frequency for resonator with an increase in intracavity photon numbers at different applied fields. Here $\Delta = \omega_{s} -\omega_0$, where $\omega\rm_{s}$ is the resonant frequency of the circuit with intracavity photon number $n\rm_{c}$. The experimental data are plotted with scatter points and fits are shown by solid lines. The obtained fits represent good agreement with the observed detuning of the resonant frequency.}
    \label{fig:Fig1}
\end{figure}

We first measured the resonator (5.6268$\,$GHz) with a weak probe and a strong blue-detuned pump tone to estimate the Kerr factor. For a Kerr nonlinear oscillator coupled to a symmetric feedline (transmission line) geometry, the equation of motion (EOM) for the intracavity field amplitude $A$ can be described using standard input-output theory \cite{yurkebuks2006}:
\begin{equation}
    \dot{A} = \left[i(\omega_0 + K|A|^2) - \frac{\kappa}{2}\right]A 
    + i\sqrt{\frac{\kappa_{\rm{ext}}}{2}}\,S_{\rm in},
    \label{eq:EOM}
\end{equation}
where $\kappa = \kappa_{\rm{ext}} + \kappa_{\rm{int}}$ is the total linewidth, and $\kappa_{\rm{ext}}$ is the external linewidth of the resonator. Since the feedline supports bidirectional propagation, the input drive $S_{\rm in}$ couples to the resonator at half the total external decay rate, giving the $\sqrt{\kappa_{\rm ext}/2}$ prefactor in the above equation. This EOM can be studied under a strong pump tone combined with a weak probe tone to estimate the Kerr factor \cite{fanisani2021levelattraction, bothner2024niobium}. The resonator was probed with $P\rm_{pr}$ = -125$\,$dBm on the chip and a blue-detuned pump signal supplied 50$\,$kHz away from resonance. To estimate the Kerr factor via two-tone spectroscopy the pump power ($P\rm_{p}$) was swept from -80 to -65 dBm as shown in Fig. \ref{fig:Fig1}(a). In this measurement no systematic variation in the internal linewidth was observed. This means we do not need to include the power dependence of TLS and two photon loss in $\kappa\rm_{int}$. The steady state solution of Eq. \ref{eq:EOM} under a strong pump tone gives \cite{fanisani2021levelattraction},
\begin{equation}
    K^2 n_c^3 - 2\Delta_{\rm p} K n_c^2 + \left(\Delta_{\rm p}^2 + 
    \frac{\kappa^2}{4}\right) n_c - \frac{\kappa_{\rm ext}\,P_{\rm p}}
    {2\hbar\omega_{\rm p}} = 0
    \label{eq:cubic}
\end{equation}
where $\Delta_{\rm p} = \omega_{\rm p} - \omega_0$, $\omega_{\rm p}$ is the angular pump frequency, $P_{\rm p}$ is the on-chip pump power in watts, and $n\rm_{c}$ is the intracavity photon number. As the steady state is set by the strong pump tone, one can linearize around the steady state solution to obtain the response of a small probe tone \cite{yurkebuks2006,fanisani2021levelattraction,bothner2024niobium}. Analyzing the EOM under the two tone gives rise to the following linearised signal and idler EOM matrix (derived in the Appendix):
\begin{equation}
    \rm{\textbf{M}} = \begin{pmatrix}
        -\dfrac{\kappa}{2} - i\Delta_{\rm eff} & iKn_c \\[8pt]
        -iKn_c & -\dfrac{\kappa}{2} + i\Delta_{\rm eff}
    \end{pmatrix}
    \label{eq:M}
\end{equation}
where $\Delta_{\rm eff} = \Delta_{\rm p} - 2Kn_{\rm{c}}$. The eigenfrequencies of the coupled signal-idler system are found by setting $\det\left(i\Omega\,\mathbb{\textbf{I}} - \rm{\textbf{M}}\right) = 0$. Evaluating this determinant yields a quadratic equation in the signal offset $\Omega$, which gives the signal frequency ($\omega\rm_{s}$) and idler frequency ($\omega\rm_{i}$) as:
\begin{equation}
    \omega_{i/s} = \omega_{\rm p} + i\frac{\kappa}{2} 
    \pm\sqrt{(\Delta_{\rm p} - Kn_c)(\Delta_{\rm p} - 3Kn_c)}
    \label{eq:SignalIdler1}
\end{equation}
where $k$ is the linewidth of the signal and idler modes. As the Kerr factor is negative, we use the smaller solution of Eq. \ref{eq:SignalIdler1} to fit for the change in resonant frequency in Fig. \ref{fig:Fig1}(a). 
For the analysis to hold, we require $P\rm{_p}$ $>>$ $P\rm_{pr}$ which was enforced by a minimum 45$\,$dBm power difference between pump and probe. We fit each vertical slice in Fig. \ref{fig:Fig1}(a) using the circle fit method \cite{probst2015resonator}, to obtain the resonant frequency ($f\rm_0$), internal quality factor ($Q\rm_{int}$), external quality factor ($Q\rm_{ext}$), and total quality factor ($Q$). Eq. \ref{eq:cubic} and \ref{eq:SignalIdler1} are then solved recursively to obtain the fits shown in Fig. \ref{fig:Fig1}(b) and a precise estimate of the Kerr factor. Here an initial guess for $n\rm_{c}$ is made using $n\rm_{c} =(\textit{Q}^2\textit{P}\rm_{p})/(\pi\hbar \textit{f}\rm_{0}^2\textit{Q}\rm_{ext})$ \cite{geaney2019nsmm} which is then substituted into the Eq. \ref{eq:cubic} to obtain an initial estimate of the Kerr factor, and an initial estimate of  $\omega\rm_{s}$ from Eq. \ref{eq:SignalIdler1}. This is then compared with the experimental data, and the estimates refined in an iterative loop. The final fit showing a zero-field Kerr factor of -2.14$\,$mHz/photon is shown in Fig. \ref{fig:Fig1}(b) along with the Kerr factors estimated for higher magnetic fields. The extracted zero-field Kerr factor can be compared to an estimate using a simplified expression developed for a  lumped-element $LC$ circuit terminated with a DC SQUID: $K = -(e^2 p^3)/(2\hbar C) $, where $e$ is the electronic charge and $p = (2\textit{L}\rm_{J}/({\textit{L}+\textit{L}\rm_{J}}))$ is the participation ratio \cite{fanisani2021levelattraction}. For our transmission line circuit we calculate the equivalent lumped-element values of $L$ and $C$, and use an $L\rm_{J}$ value corresponding to a critical current of 330$\,\mu$A previously determined \cite{bhandari2026nbnanobridge}. This gives a Kerr factor of -2.13$\,$mHz/photon which is in close agreement with the experimentally obtained value. Beyond an applied field of 2.48$\,$mT (the highest value shown in Fig. \ref{fig:Fig1}(b) we observed that the resonant frequency shows a jump rather than a smooth decrease with an increase in pump power. We attribute this behaviour to the movement of flux vortices in and out of the resonator on application of a strong pump signal.
\subsection{Red-detuned Regime}
\begin{figure}[!t]
    \centering
\includegraphics[width=\columnwidth]{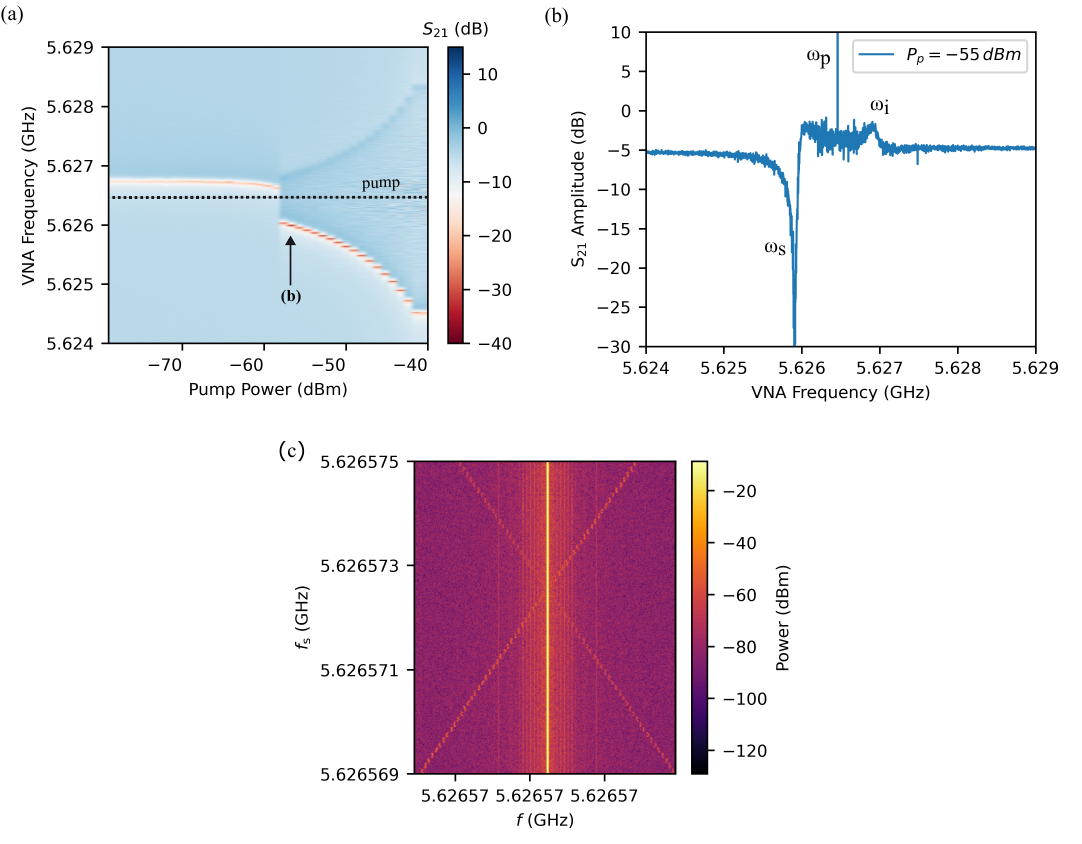}
    \caption{(a) A colour plot showing the transmission spectrum of resonator with an increase in pump power for a red-detuned signal. The resonant frequency shows an initial slight shift as pump power increases and then a jump below the pump frequency and emergence of net idler gain. (b) A line plot showing the resonant dip and the idler gain for $P\rm_{p} =-55\,$dBm. (c) A colour plot showing the compiled results from the spectrum analyser measurements, where the pump tone was fixed (vertical line), and the idler frequency (negative slope line) changes according to the degenerate four wave mixing process as the signal tone (positive slope line) is swept.}
    \label{fig:fig2}
\end{figure}

As the Kerr factor for the circuit is negative, we choose a pump frequency below the resonant frequency in order to pump higher number of photons into the resonator to drive it beyond bifurcation to study idler gain. The results shown in Fig. \ref{fig:fig2}(a) are plotted for a slightly higher probe power (-115$\,$dBm) then used in the blue-detuned measurements to show a clearer indication of idler gain to the reader. A line plot of the transmission spectrum obtained for $P\rm_{p} =-55\,$dBm and $\omega\rm_{p}/(2\pi) =$ 5.62645$\,$GHz is shown in Fig. \ref{fig:fig2}(b), where the dip corresponds to the signal and the peak corresponds to the idler gain. The VNA trace in Fig. \ref{fig:fig2}(b) shows a clear four-wave-mixing response with $\omega\rm_{s}+\omega\rm_{i} = 2\omega\rm_{p}$. The generation of the idler was also confirmed in a separate experiment by mixing two tones and observing the emergence of a third tone using a spectrum analyzer, as shown in Fig. \ref{fig:fig2}(c). In this experiment, a small signal tone at variable frequency was generated using a Windfreak synthesiser and a strong pump tone of -68$\,$dBm at a frequency of 5.62645$\,$GHz was generated using a Keysight signal generator. These were added using a directional coupler. The line with a negative slope in Fig. \ref{fig:fig2}(c) shows how the idler frequency shifts as the probe (signal) frequency is swept in agreement with $\omega\rm_{s}+\omega\rm_{i} = 2\omega\rm_{p}$. No idler was detected below a pump power of -68$\,$dBm which is likely set by the noise floor of the measurement.
\label{Sec:Red-detuned Regime}

\subsection{Demonstration of Parametric Gain}
\label{Sec:Demonstration of Parametric Gain}
\begin{figure}[!t]
    \centering
\includegraphics[width=\columnwidth]{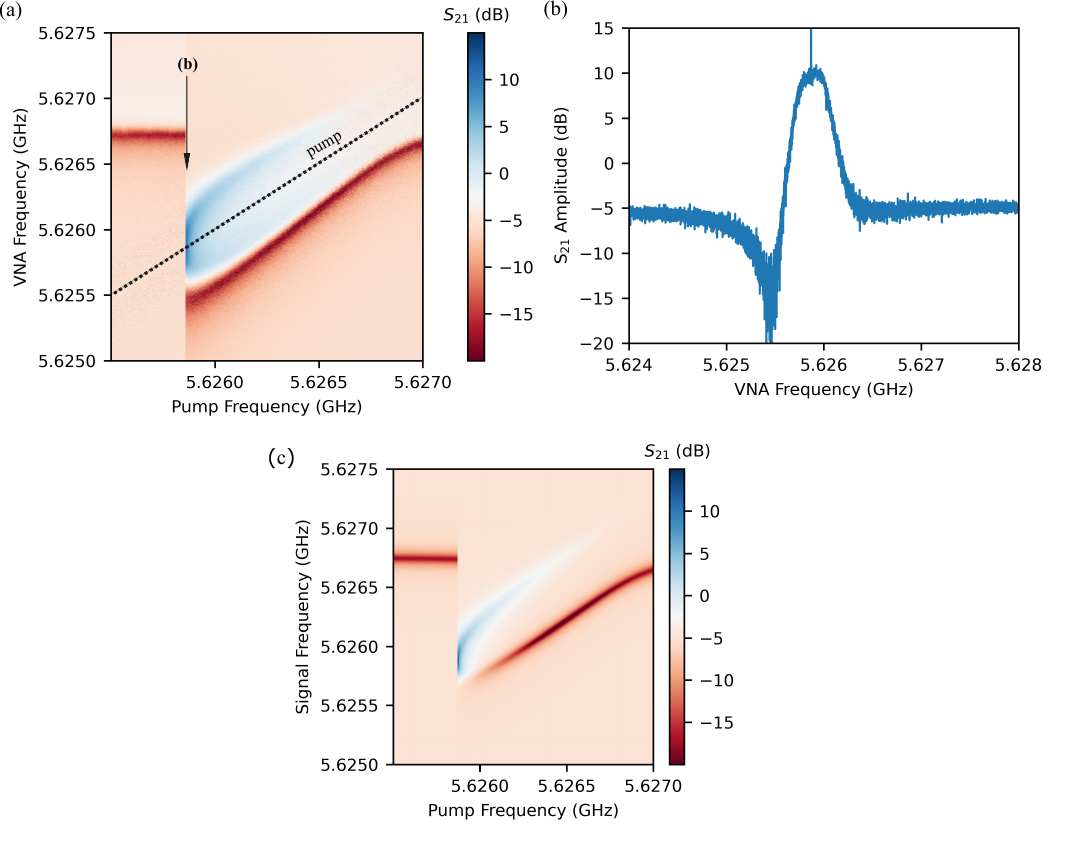}
    \caption{(a) A colour plot showing the measured transmission spectrum for resonator as the pump frequency is decreased in steps of 6$\,$kHz starting from the right-hand side of the plot. As the pump frequency is decreased the idler appears and its frequency (peak) tracks the signal frequency (dip) until a maximum gain is reached.  Beyond that point the resonator snaps back to a single resonant dip. (b) A line plot showing the maximum gain of 15$\,$dB (with respect to the baseline at -5$\,$dB) with a bandwidth of 300$\,$kHz. (c) The solution of the analytical model discussed in the main text for $\omega\rm_{0}/2\pi=5.62675\,$GHz, $\kappa\rm_{int}=38\,$kHz, $\kappa\rm_{ext}=120\,$kHz, $P\rm_{p} =-60\,$dBm, and $K=-2.14\,$mHz/photon. The solution has been offset by -5$\,$dB to match the baseline of the experimental data.}
    \label{fig:fig3}
\end{figure}
To search for a region of maximum parametric gain, we follow a similar procedure to that in \cite{fanisani2021levelattraction} for observing gain in a high-\textit{Q}/low-\textit{K} devices. Beyond the bifurcation threshold, the nonlinear resonator (Eq. \ref{eq:cubic}) has two locally stable solutions corresponding to high and low values of $n\rm_{c}$, whereas the third solution is unstable. The high $n\rm_{c}$ branch corresponds to higher amplitude pump current and corresponds to the right-hand side of Fig. \ref{fig:fig2}(a) where both the signal dip and the idler peak are present. To observe parametric gain in high-\textit{Q}/low-\textit{K} devices requires sitting on the higher $n\rm_{c}$ branch. To reach the higher $n\rm_{c}$ branch we start from the lower branch by setting $P\rm_{p}=-60\,$dBm  and $\omega\rm_{p}/2\pi=5.627\,$GHz. We then gradually reduce the pump frequency $\omega\rm_{p}/2\pi$ in 6$\,$kHz steps (adiabatically) as shown in Fig. \ref{fig:fig3}(a). Initially $S\rm_{21}$ shows just a dip (on the right-hand side of Fig. \ref{fig:fig3}(a), but as $\omega\rm_{p}$ is reduced further we observe that the resonator switches to the higher $n\rm{_c}$ branch as confirmed by the presence of both the signal dip and idler peak. As $\omega\rm_{p}$ is decreased further the signal and idler frequencies track each other down, with an increase in the linewidth of the two modes. A further decrease in $\omega\rm_{p}$ leads to an increase in the net gain until a maximum gain of 15$\,$dB (estimated from the baseline at -5$\,$dB) is reached at the position indicated by an arrow on Fig. \ref{fig:fig3}(a). The maximum gain line-shape has a bandwidth of $\simeq$ 300$\,$kHz as shown in Fig. \ref{fig:fig3}(b), where the line in the middle of the gain peak is the pump tone. A further decrease in the pump frequency causes the resonator to switch to the lower branch with only the dip at the bare resonant frequency of the circuit present, as shown on the left-hand side of Fig. \ref{fig:fig3}(a).\\
To understand the full behaviour in Fig. \ref{fig:fig3}(a), we solved the Eq. \ref{eq:cubic} using the resonant frequency, the internal/external linewidths obtained for a probe power of -125$\,$dBm, the zero-field Kerr factor determined previously in Section \ref{sec:Blue-detuned Regime}, and the experimentally set pump power. Here, we specifically selected the highest solution for $n\rm_{c}$ in Eq. \ref{eq:cubic}, The EOM Matrix \textbf{M} for the solution at signal frequency $\Omega$ can be related to the measured $S\rm_{21}$ at signal frequency $\Omega$ using input-output theory as
\begin{equation}
    S_{21,\mathrm{dB}}(\Omega)
    =
    20\log_{10}
    \left|
    \left[{\mathbb{\textbf{I}}-\frac{\kappa_e}{2}\left(i\Omega\mathbb{\textbf{I}}-\textbf{M}\right)^{-1}}\right]_{00}
    \right|.
    \label{eq:S21_dB}
\end{equation}
Using this relation, the full analytical solution is shown in Fig. \ref{fig:fig3}(c) offset by -5$\,$dB to match the baseline of the experimental data. The analytical solution and experimental data are in good agreement with the same qualitative features of tracking (interlocking) of the signal and idler modes, increase in idler gain, and then a sudden snapping to the lower branch with a dip at $\omega_0$.
\section{Conclusion}
\label{Sec:Conclusion}
In this study we have explored the non-linear behaviour  at 15$\,$mK of a Nb nanobridge DC SQUID embedded in a CPW resonator and estimated the Kerr factor up to a perpendicular applied field of 2.48$\,$mT. We showed the resonator can be operated as a parametric amplifier in the 4-wave-mixing regime providing a maximum gain of 15$\,$dB. This opens up the possibility of studying the noise and dynamic range of such amplifiers, and also the further possibility of developing 3-wave-mixing JPAs based on Nb nanobridges. The device's fairly high critical current and hence small Kerr factor, allows for its implementation as a high saturation power JPA \cite{naaman2017highsat,white2023highdynamic}. Four-wave mixing would allow noise reduction through quantum mechanical squeezing, further enhancing the amplifier performance. The demonstration that the device behaviour is well-understood using the Kerr nonlinear oscillator model suggests one can further improve the gain and bandwidth by using similar methods to those employed for JPAs, for instance by increasing the coupling capacitance to put the resonator into a highly over-coupled regime with an overall low-Q \cite{castellanos2007parametric,butseraen2022graphene}. A higher Kerr factor could be achieved by fabricating variable thickness (3D-like) nanobridges, or by increasing the operating temperature. With application-specific improvements, the device has the potential of being deployed as a qubit readout \cite{vijay2011quantumjumps,lin2013singleshot} exploiting its high saturation power, or in the amplification of ultra-weak microwave signals in fundamental physics experiments \cite{amad2025neutrino} exploiting its magnetic field resilience. 
\section*{Acknowledgments}
This work was supported by the UK Science and Technologies Funding Council (STFC) via grants ST/T006064/1, ST/T006137/1 \& ST/T006099/1, by the UK National Quantum Technologies Programme, and by an EPSRC/NPL co-funded PhD Studentship at UCL.
\section*{Appendix: Linearised Signal-Idler Equations of Motion}
\label{sec:AppendixB}
To solve the full two-tone EOM, we move to a frame rotating at the pump frequency and expand the internal field using the ansatz
\begin{equation}
    A = \left(\alpha_0 + \delta_- e^{i\Omega t} 
    + \delta_+ e^{-i\Omega t}\right)e^{i\omega_{\rm p}t},
    \label{eq:ansatz}
\end{equation}
where $\delta_-$ and $\delta_+$ are the small classical signal and idler amplitudes at offset $\Omega = \omega_{\rm pr} - \omega_{\rm p}$ from the pump. Both sidebands must be retained because the Kerr nonlinearity $iK|A|^2A$ inherently couples the positive frequency component ($e^{i\Omega t}$) with the negative frequency component ($e^{-i\Omega t}$). By linearising the equation of motion around the steady-state pump solution (setting $\alpha_0$ to be real such that $\alpha_0^2 = n_c$), we obtain a set of coupled, single-line linear equations for the signal and conjugated idler:
\begin{align}
    i\Omega\,\delta_- &= \left(-\frac{\kappa}{2} 
    - i\Delta_{\rm eff}\right)\delta_- + iKn_c\,\delta_+^*, 
    \label{eq:delta_minus}\\
    i\Omega\,\delta_+^* &= \left(-\frac{\kappa}{2} 
    + i\Delta_{\rm eff}\right)\delta_+^* - iKn_c\,\delta_-.
    \label{eq:delta_plus}
\end{align}
In matrix form, Eqs \eqref{eq:delta_minus}-\eqref{eq:delta_plus} read
\begin{equation}
    i\Omega\begin{pmatrix}\delta_- \\ \delta_+^*\end{pmatrix}
    = \textbf{M} \begin{pmatrix}\delta_- \\ \delta_+^*\end{pmatrix},
    \label{eq:matrix_EOM}
\end{equation}
where the internal equations-of-motion matrix is purely a function of cavity parameters and steady-state population:
\begin{equation}
    \textbf{M} = \begin{pmatrix}
        -\dfrac{\kappa}{2} - i\Delta_{\rm eff} & iKn_c \\[8pt]
        -iKn_c & -\dfrac{\kappa}{2} + i\Delta_{\rm eff}
    \end{pmatrix}.
\end{equation}
\bibliographystyle{unsrt}
\bibliography{references}
\end{document}